\documentclass[
aps,
pre,
amsmath,amssymb,
reprint,
showkeys,floatfix
]{revtex4-1}

\usepackage[T1]{fontenc}
\usepackage[utf8]{inputenc}
\usepackage{amsmath}
\usepackage{amssymb}
\usepackage{graphicx}
\usepackage{skak}
\usepackage{multirow}
\usepackage{epsdice}
\usepackage{soul}
\usepackage[
citecolor=blue,
colorlinks,
linkcolor=blue,
urlcolor=blue,
]{hyperref}
\usepackage{epstopdf}
\usepackage{bm}
\usepackage{braket}
\usepackage[version=4]{mhchem}
\usepackage{tikz}
\usetikzlibrary{arrows,shapes,trees}
\usepackage{lipsum} 

\begin{document}
	\title{Coordinating cooperation in stag-hunt game: Emergence of evolutionarily stable procedural rationality}
	
	\author{Joy Das Bairagya}
	\email{joydas@iitk.ac.in (Corresponding author)}
	\affiliation{
		Department of Physics,
		Indian Institute of Technology Kanpur, Uttar Pradesh, PIN: 208016, India
	}
	\author{Sagar Chakraborty}
	\email{sagarc@iitk.ac.in}
	\affiliation{
		Department of Physics,
		Indian Institute of Technology Kanpur, Uttar Pradesh, PIN: 208016, India
	}

\begin{abstract}
Humans are bounded rational at best and this, we argue, has worked in their favour in the hunter-gatherer society where emergence of a coordinated action, leading to cooperation, is otherwise the standard stag-hunt dilemma (when individuals are rational). In line with the fact the humans strive for developing self-reputation by having less propensity to cheat than to be cheated, we observe that the payoff structure of the stag-hunt game appropriately modifies to that of coordination-II game. Subsequently, within the paradigm of evolutionary game theory, we establish that a population---consisting of procedural rational players (a type of bounded rationality)---is unequivocally evolutionarily stable against emergence of more rational strategies in coordination-II game. The cooperation is, thus, shown to have been established by evolutionary forces picking less rational individuals. 
\end{abstract}

\maketitle
\section{Introduction}
Coordinated actions among living being are essential for survival. Efficient choice of new nests in social insects~\cite{Seeley1999,Visscher2007,Visscher-kirk2007,Franks2008}, communal routes of migratory birds~\cite{Wallraff1978,SIMONS2004,Biro2006}, collective selection of bats' roosting sites~\cite{Kerth2006}, swarms of insects~\cite{Buhl2006}, shoals of fishes~\cite{Reebs2000,Hemelrijk2008,Ward2008}, flocks of birds~\cite{shapiro1999,Ballerini2008}, group hunting~\cite{Gompper1996,Boesch1989,Courchamp2002}, herds of ungulates~\cite{Gueron1996,Conradt1998,RUCKSTUHL1998,FISCHHOFF2007825,GAUTRAIS20071443} and a few troops of primates~\cite{Stewart1994,Trillmich2004,Meunier2004,Sellers2007,SUEUR200884}
 are few such examples. Likewise, coordinated actions are paramount for modern human societies, e.g., using a common language during conversation, conventions of car driving on the left side or right side of the road, heavily concentrated industries in a small area, using same measuring convention~\cite{camerer2011behavioral}. The tendency of coordination is not limited to modern human societies, it could be traced back to our ancestors forming hunter-gatherer societies where coordinated efforts must have been present in the phenomena of social status levelling~\cite{wiessner1996food,boehm2009hierarchy}, non-kin food sharing~\cite{Gurven_2004}, cooperative food acquisition~\cite{Hill2002}, provisioning of multiple goods and services~\cite{Gurven_2004} and alloparental caretaking~\cite{hewlett2017hunter}. Coordination is somewhat special in humans, as it is documented even between genetically unrelated individuals; it has been hypothesized to be one of the primary evolutionary forces that favoured \emph{Homo sapiens} against other hominid species in the Pleistocene epoch and became dominant among our species in the Holocene~\cite{Hill2009}.
 
 There are a plethora of studies to comprehend the emergence of the coordination between genetically non-related individuals~\cite{Nowak2006Coop}. One of the most useful models to mathematize social coordination is {game theory}. Cooperation, {manifested as a coordinated action}, in hunter-gatherer societies or any real-life situation can be succinctly represented through coordination games~\cite{skyrms2004stag}, like the stag-hunt game, which mathematize the conflict of interests---whether form a group to hunt a stag or to settle for less payoff by hunting a hare alone. The payoff matrix
 \begin{table}[h!]
	\centering
	\begin{tabular}{cc|c|c|}
		& \multicolumn{1}{c}{} & \multicolumn{2}{c}{{Player-$2$}}\\
		& \multicolumn{1}{c}{} & \multicolumn{1}{c}{C} & \multicolumn{1}{c}{\,\,\,\,${\rm D}$\,\,\,\,}\\\cline{3-4} 
		\multirow{2}*{{Player-$1$}} & C & $1,1$ & $S,T$ \\\cline{3-4}
		& ${{\rm D }}$ & $T,S$ & $0,0$ \\\cline{3-4} 
	\end{tabular}\quad
\end{table}
\\
can { succinctly represent interaction} between two individuals who can either cooperate (C) for getting higher payoff or  they defect (D) with each other, accepting a lower payoff. In the payoff matrix, the first and the second element of each cells are the payoffs of Player-1 and Player-2, respectively. Additionally, to represent a coordination game, one fixes $1>T$ and $0>S$; specifically, for a stag-hunt game, the ordinal relationship between payoffs is  {$1>T\geq0>S$}. 

It is interesting to know that evolution has shaped humans to be extremely sensitive towards cheater-detection mechanism that has not been found in chimpanzees and other greater apes~{\cite{Cosmides1989}}: Humans are much concerned about their self-reputation and would not want to be seen as a cheater~{\cite{Tomasello2012,Milinski2002}}; they would rather prefer to be cheated. This feature inhibits the tendency of free riding and promotes cooperation~\cite{Tomasello2012}. So, in the earliest manifestation of human collaborations (foraging in gatherer-hunter society), the payoff of non-collaboration when others are ready to collaborate is evaluated lesser than that of the individual who is always ready to collaborate.  Hence, for gatherer-hunter individuals the ordinal relation of the stag-hunt game should modified to $1>0>S>T$: In literature~\cite{Hummert2014}, this game is known as coordination-II game.

\section{The question}
Howsoever simple this game may appear, theoretically however, it is not unanimously clear how hunter-gatherer (even with the coordination-II game) could resolve the conflict between two NE outcomes. Neo-classical economic theory presumes human behaviour to be as consistent (or rational) { over available preferences} as it could be. The desire of human beings to opt for one action over another can be mathematized by assuming that there are some utilities associated with the actions and everyone tries to maximize the utility. Through the celebrated von Neumann--Morgenstern (VNM) utility theorem~\cite{VNMbook}, one ensures the existence of expected utility that in a consistent way expresses the order of preferences over various events.  For any two-player game between such VNM rational (VR) players, one can mathematically show if everyone has mutual knowledge about the set of players, the sets of actions, the payoff functions, and everyone's true belief about the other player, then their beliefs {lead to} an equilibrium, which is Nash equilibrium (NE)~\cite{Brandenburger1992}---a strategy profile, unilateral deviation from which does not fetch any additional payoff to the deviating player.  Since either  cooperation or defection can emerge as a  NE in a coordination game, the stag-hunt dilemma arises. 

A moment of reflection tells us that it is pretty evident that humans, indeed, do not possess infinite mental capabilities. Nevertheless, it is also not true that humans are incredibly foolish. The truth lies in between these two extremes.  Therefore, one must stray from the assumptions of rationality and strict requirement of mutual (or common) knowledge towards bounded rationality to comprehend human decision making process in the gatherer-hunter societies.  Indeed, for modern humans it has been experimentally observed that humans do not act VNM rationally in every situation~\cite {Busemeyer1993,Rieskamp2006,Selten2008}. In fact, such considerations gave birth to modern behavioural economics~\cite{camerer2011behavioral}. Specifically, as empirical observation~\cite{Selten2008} suggests, human behaviour in a strategic interaction could be explained more accurately through the payoff sampling equilibrium \textcolor{black}{(SE($k$)---$k$ in the notation will become clear in due course)} than the Nash equilibrium. Sampling equilibrium is reached by procedurally rational---a kind of bounded rational---players who literally or virtually  sample  each of their actions  and subsequently, choose the respective actions that yield the best outcomes during sampling process.


Up to now we have only pondered over the possible equilibrium outcomes form a rather static-approach: the dynamics of a decision-process leading to the corresponding equilibrium has not been considered. It is well-known that the best-response dynamics~\cite{Milgrom1990,Matsui1992} leads to NE outcome; whereas, the sampling dynamics~\cite{Sethi2000} yields SE(1) outcome asymptotically. In the specific case of coordination-II game, both C and ${\rm D}$ are both NE and SE(1); however, while both are reached under best-response dynamics, only C is reached by all initial beliefs (except for the one corresponding to ${\rm D}$)~\cite{Sethi2000} (see also Appendix~\ref{app:A}).  Hence, it appears that a deviation from rationality could explain the attainment of cooperation. However, it remains to be seen whether such a deviation emerges and sustains itself evolutionarily---after all, there can always be appearance of individuals (mutants) with alternate level of rationality.

Such a question automatically leads one to consider the paradigm of evolutionary game theory~\cite{smith1982evolution}. In this backdrop, rationality can be seen as a behavioural trait~\cite{Houston2007,Waksberg2008,Brennan2018} of humans and, naturally, it is amenable to Darwinian evolution. Various degrees of rationality may be seen as various values of the trait. Thus, the technical question is whether in a population of procedurally rational players playing the modified version of stag-hunt game (viz., coordination-II), some VNM rational mutants can invade the population; in other words, whether procedural rationality is evolutionarily stable strategy (ESS)~\cite{Taylor1978}? Consequently, we need to closely scrutinize the scenario where a VNM rational player (henceforth, VR-player) and a procedurally rational player (henceforth, {PR($k$)}-player) strategically interact. Specifically, we need to determine if { an equilibrium in belief is asymptomatically reached}  by the pair of players and, thence, how much payoffs are received by them. 

\section{Setup}
To this end, recall that in a one shot two-player non-cooperative game, it is unfeasible for a player to know her opponent's strategy; she can at best have a belief, ${\bm x}=(x,1-x)$---a probability distribution over the action set, $\{{\rm C},{\rm D}\}$, of her opponent. Given a belief, a player would choose best response (strategy) to it. However, the player knows that the opponent would do the same and would like to take that into account as well and thus, would update her belief accordingly. Since opponent would have similar thoughts, the player knows that the opponent knows that she would update her belief and hence, opponent would tune her belief too; thus, the player would update her belief further. This mental process would go on till an equilibrium in belief is reached, if at all. One can mathematize this virtual experimentation of belief-updating as given below where one can note that the association of probability distribution over the consequences of a player's actions~\cite{Osborne1998} in the {virtual experimentation} view of sampling equilibrium is assumed to have been extracted from her belief about the opponent.
Let us assume that after $m$-times updating, the belief of Player-1, i.e., the probability with which Player-2 chooses C, becomes $x_m$. [Since there are only two actions, belief, ${\bm x}=(x,1-x)$, can be completely specified by only one component, say, $x$.]  Similarly, let Player-2's belief about Player-1 be $y_m$---the probability with which Player-1 chooses  C. Thus, Player-1 assumes at $m$th step that C { would have been} played by Player-2 $mx_m$ times and at $(m+1)$-th step whether C is played by Player-2 depends on her best response to $y_m$ (Player-2's belief). Similarly and simultaneously, Player-2 assumes at $m$th step that C would be played by Player-1 $my_m$ times and at $(m+1)$-th step whether C is played by Player-1 depends on her best response to $x_m$. Mathematically,
\begin{subequations}
	\label{eq:belief_update}
	\begin{eqnarray}
		x_{m+1}&=&\frac{m}{m+1}x_{m}+\frac{1}{m+1}{\rm BR}(y_m),\label{update_ply_1}\qquad\\
		y_{m+1}&=&\frac{m}{m+1}y_{m}+\frac{1}{m+1}{\rm BR}(x_m);\label{update_ply_2}
	\end{eqnarray}
\end{subequations}
where ${U}_ {\rm C}$ and ${U}_{{\rm D}}$ are allowed to, respectively, denote the utilities corresponding to playing actions ${\rm C}$ and ${{\rm D}}$. Consequently, ${\rm BR}(x)$ function takes values $1$, $0$ and $1/2$ for {\color{black}PR($k$)}-player or $x$ for VR-player if for belief $x$, ${U}_ {\rm C}> {U}_{{\rm D}}$, ${U}_ {\rm C}< {U}_{{\rm D}}$ and ${U}_ {\rm C}= { U}_{{\rm D}}$, respectively. Note that for the above update of beliefs to be mentally done as virtual experiments,  the types of players, update rules, and payoff matrix should be mutual knowledge. While initial beliefs should also be mutual knowledge but that is not a necessary requirement for our later results because they will be presented after taking average over all possible initial beliefs.

While the above model of belief updating remind one of the fictitious play~\cite{Gaunersdorfer1995}, it is crucial to realize that meaning of utility for a VR-player and a PR($k$)-player is not same. Let us elaborate on this:
\begin{itemize} 
	
\item {VR-player's utility:} She calculates utility as the expected payoff in line with the traditional meaning of VNM rationality. Therefore, given the beilef is $x$ about the opponent's strategy, ${U}_{\rm C}\equiv x\times 1+(1-x)\times S$ and ${U}_{{\rm D}}\equiv x\times T + (1-x)\times 0$. Hence, the utilities of the PR($\infty$)-player and the VR-player are identical.

\item {{\color{black}PR(1)}-player's utility:} In this case, the player mentally samples the outcome once for each action, C and ${\rm D}$, then she decides on the utilities according to the outcomes corresponding to her sampled actions.  Virtual experimentation in her mind simulates the sampling process as  follows: She flips a biased coin in which probability of `heads' is $x$ (same as her belief). She determines the outcomes for each of her two actions by flipping the coin once for each action independently. Then the utility of the action C, i.e., ${U}_{\rm C}$  is $1$ or $S$ depending on whether she gets a `heads' or a `tails'. Similarly, the utility of the action ${\rm D}$, i.e., ${ U}_{{\rm D}}$ is $T$ or $0$ depending on whether she gets a `heads' or a `tails'. Thus, the utilities for actions C and ${\rm D}$ are, respectively, two random numbers ${U}_{\rm C}\in \{1,S\}$ and ${U}_{{\rm D}}\in \{T,0\}$. 

{\color{black}\item {PR($k$)-player's utility:} Similar to the PR(1)-player, first she mentally samples the outcome $k$ times for each action, C and ${\rm D}$; then she determines the utilities in line with the outcomes corresponding to her sampled actions.  Virtual experimentation goes as  follows: She conducts tossing experiment with the aforementioned biased coin whose `heads' corresponds to C and the `tails' corresponds to D. She determines the sequence of outcomes for each of her two actions by flipping the coin $k$ many times for each action independently. The utility of the action C is defined as ${U}_{\rm C} = \frac{1}{k}\sum_{i=1}^k \omega^{\rm C}_i$, where $\omega^C_i$  is $1$ or $S$ depending on whether she gets a `heads' or a `tails' in $i$th flip. Similarly, the utility of the action ${\rm D}$ is ${ U}_{{\rm D}} = \frac{1}{k}\sum_{i=1}^k \omega^D_i$, where $\omega^{\rm D}_i$  is $T$ or $0$ depending on whether she gets a head or tail  in $i$th flip. Thus, the utilities for actions C and D are, respectively, two random numbers ${U}_{\rm C}$ and ${U}_{{\rm D}}$. 

While, of course, setting $k=1$ gives back the utilities written earlier for the PR(1)-player, an important comment about the case $k\to\infty$ is in order: Note that by the law of large numbers, it clearly follows that
\begin{align*}
	U_{\rm C} &= \lim_{k \to \infty} \frac{1}{k} \sum_{i=1}^k \omega^{\rm C}_i = x \times 1 + (1 - x) \times S, \\
	U_{\rm D} &= \lim_{k \to \infty} \frac{1}{k} \sum_{i=1}^k \omega^{\rm D}_i = x \times T + (1 - x) \times 0.
\end{align*}
Interestingly, the utilities obtained by PR($\infty$)-player is nothing but the expected utilities as would be estimated by a VR-player. Consequently, wherever appropriate, we use PR($\infty$) and VR  synonymously, although it should always be borne in mind that conceptually both the players use fundamentally different processes for estimating the utilities.
}
\end{itemize}
It is worth remarking here that if we recall that one of the perspectives of looking at rationality is `consistency among preferences'  (i.e., preferences follows the axioms~\cite{VNMbook} of completeness, transitivity, continuity and independence), then {\color{black}PR(1)}-player in not strictly rational because it doesn't respect the continuity and independence axioms---a fact manifested through random nature of  her utility.

{ Suppose, asymptotically, all individuals go towards some stationary beliefs which do not change further under Eq.~\ref{eq:belief_update}. This stationary belief profile is `equilibrium in belief': The equilibrium strategy $p$ played by Player-1 is Player-2's equilibrium belief, $\lim_{m\to\infty}y_m=y_{\infty}$ (say) and the equilibrium strategy $q$ played by Player-2 is Player-1's equilibrium belief, $\lim_{m\to\infty}x_m=x_{\infty}$ (say). Of course, we expect that if both players are VR-players, then $(p,q)$ is NE and if both players are {\color{black}PR(1)}-players, then $(p,q)$ is SE(1). However, it is not obvious what $(p,q)$---equivalently, $(y_\infty,x_\infty)$---corresponds to if VR-player plays with {\color{black}PR(1)}-player. In general, for payoff matrix ${\sf A}\equiv\left[^1_T~^S_0\right]$, the average payoffs of Player-1 and Player-2, respectively, are ${\bm y}_{\infty}\cdot{\sf A}{\bm x}_{\infty}$  and ${\bm x}_{\infty}\cdot{\sf A}{\bm y}_{\infty}$, respectively, where ${\bm x}_{\infty}=\left({x_{\infty}},{1-x_{\infty}}\right)$ and ${\bm y}_{\infty}=\left({y_{\infty}},{1-y_{\infty}}\right)$ are two column vectors.}

In coordination-II game, as far as the case of two strategically interacting {\color{black}PR(1)}-players are considered, the future of initial beliefs under Eq.~(\ref{eq:belief_update}) is rather straightforward: The asymptotic beliefs of both the players always reach $(x_{\infty},y_{\infty})=(1,1)$, except if $(x_0,y_0)=(0,0)$. The case of two VR-players interacting is little more interesting. Owing to the ordinal relationship $1>0>S>T$, the symmetric mixed NE, $|S|/(1+|T|+|S|)$, always remains less than $1/2$; additionally, there exist two pure NEs: $(0,0)$ and $(1,1)$.  Obviously, if $x_0,y_0<|S|/(1+|T|+|S|)$, then under Eq.~(\ref{eq:belief_update}), $(x_\infty,y_\infty)\to(0,0)$; similarly, if $x_0,y_0>|S|/(1+|T|+|S|)$, then $(x_\infty,y_\infty)\to(1,1)$. 

However, if neither of the above conditions on initial conditions are satisfied, the asymptotic state depends on the exact shape of the basin of attractions (i.e., implicitly, on $S$ and $T$), and hence, hard to predict analytically. This analytical intractability becomes even more serious when one considers that a VR-player is interacting with a {\color{black}PR(1)}-player: The asymptotic state not only depends on the initial beliefs, it also depends on the exact values of $S$ and $T$. In general, Eqs.~(\ref{eq:belief_update}) are stochastic. So, let us scrutinize the mean-field level behaviour~{\color{black}(see also Appendix~\ref{app:comms})}. Furthermore, for mathematical simplicity, we change the difference equations to  differential equation by relabelling discrete $m$ to continuous $\tau$ and assuming that two successive steps are infinitesimally close (i.e., $\delta \tau \to 0$). The resulting equation is
\begin{subequations}
	\label{eq:belief_update_meanfield}
	\begin{eqnarray}
	\frac{d}{d\tau}\braket{x(\tau)}&\equiv&\lim\limits_{\delta \tau \to 0}\frac{\braket{x(\tau+\delta \tau)}-\braket{x(\tau)}}{\frac{\delta \tau}{m+1}}\nonumber\\
	\phantom{\frac{d}{dt}\braket{x(\tau)}}&=&{{\rm BR}(\braket{y(\tau)})}-\braket{x(\tau)},\qquad\qquad\label{update_ply_1}\\
	\frac{d}{d\tau}\braket{y(\tau)}&\equiv&\lim\limits_{\delta \tau \to 0}\frac{\braket{y(\tau+\delta \tau)}-\braket{y(\tau)}}{\frac{\delta \tau}{m+1}}\nonumber\\
\phantom{\frac{d}{dt}\braket{x(\tau)}}&=&{{\rm BR}(\braket{x(\tau)})}-\braket{y(\tau)};\qquad\qquad\label{update_ply_2}
	\end{eqnarray}
\end{subequations}
where $m+1$ has been absorbed in $d\tau$ without any loss of generality and angular brackets imply average over various realizations of time-evolution of a single initial condition, $(\braket{x_0},\braket{y_0})=(x_0,y_0)$. One should note that a crucial approximation has been made: In order to close the hierarchy of differential equations for the moments of $x$ and $y$, we have approximated $\braket{{\rm BR}({y(\tau)})}$ by ${{\rm BR}(\braket{y(\tau)})}$ and $\braket{{\rm BR}({x(\tau)})}$ by ${{\rm BR}(\braket{x(\tau)})}$. Physically, this approximation means that, as her mean-field behaviour,  each player plays best response to her average belief.

\section{Results}
 As is the template in evolutionary game theory for estimating evolutionary stability of a strategy, we consider a well-mixed unstructured population of players each of whom can act as either a VR-player or a {\color{black}PR(1)}-player. In a particular state of this population, we ask if an infinitesimal mutation would invade the resident population or not. To answer this question technically, we need to find the expected payoff (fitness) of each strategy (VR and {\color{black}PR(1)})  when they are pitted against each other.  Since every player would meet with many other players, their fitness should be seen at the mean-field level. Furthermore, since the initial beliefs can also be anything, in general, hence an additional averaging of the fitness over all possible beliefs is also appropriate. The resultant payoff matrix, ${\sf \Pi}$, of any player in the symmetric evolutionary game can be represented as follows:

\begin{table}[h!]
\hskip -1.5 cm
	\renewcommand{\arraystretch}{1.6} 
	\setlength{\tabcolsep}{14pt} 
	\begin{tabular}{cc|c|c|}
		& \multicolumn{1}{c}{} & \multicolumn{1}{c}{{\color{black}PR(1)}} & \multicolumn{1}{c}{VR} \\\cline{3-4} 
		\multirow{2}{*}{} 
		& {\color{black}PR(1)} & $\overline{\braket{{\bm y}^{\rm PR}_{\infty}} \cdot {\sf A}  \braket{{\bm x}^{\rm PR}_{\infty}}}$ 
		& $\overline{\braket{{\bm y}^{\rm VR}_{\infty}} \cdot {\sf A} \braket{{\bm x}^{\rm PR}_{\infty}}}$ \\\cline{3-4}
		& VR & $\overline{\braket{{\bm y}^{\rm PR}_{\infty}} \cdot {\sf A}  \braket{{\bm x}^{\rm VR}_{\infty}}}$ 
		& $\overline{\braket{{\bm y}^{\rm VR}_{\infty}} \cdot {\sf A}  \braket{{\bm x}^{\rm VR}_{\infty}}}$ \\\cline{3-4} 
	\end{tabular}
	\label{mat:fitness}
\end{table}
\noindent where superscripts VR and {\color{black}PR(1)} explicitly mark the player-type corresponding to the asymptotic beliefs and the bars indicate the averaging over all initial beliefs. By definition, {\color{black}PR(1)} is ESS if either
${\sf \Pi}$({\color{black}PR(1)},{\color{black}PR(1)}) $>$ ${\sf \Pi}$(VR,{\color{black}PR(1)}), or if
${\sf \Pi}$({\color{black}PR(1)},{\color{black}PR(1)}) $=$ ${\sf \Pi}$(VR,{\color{black}PR(1)}) then
${\sf \Pi}$({\color{black}PR(1)},VR) $>$ ${\sf \Pi}$(VR,VR). Similarly, it can be determined if VR is ESS.

{\color{black} We must emphasize that, thanks to the folk theorems and related results~\cite{Cressman2014} in the evolutionary game theory, we do not have to explicitly consider the dynamics of the replication-selection process leading to ESSes. For infinite population, the most widely used such dynamics is governed by replicator equation~\cite{Taylor1978, MC21} which formalizes the concept of Darwinian evolution in most simple non-trivial form. Since the fact that an ESS is an asymptotically stable fixed point of the replicator dynamics~\cite{Cressman2014}, exists as a general result, we need to only determine the ESS directly from the payoff matrix $\sf \Pi$. However, keeping in mind the replicator dynamics is necessary for having conceptual clarity about the different time-scales at play in the system. Two main time-scales are implicitly involved in our framework: one associated with the belief update dynamics, Eq.~(\ref{eq:belief_update_meanfield}), and the other with the replicator dynamics. We assume that the time required for belief equilibration is quite less compared to the time-scale of the replication-selection process: Essentially, before replication process kicks in, the players would have reached their respective belief equilibria.} 

\begin{figure}[h!]
	\centering
	\includegraphics[width=1.0\linewidth]{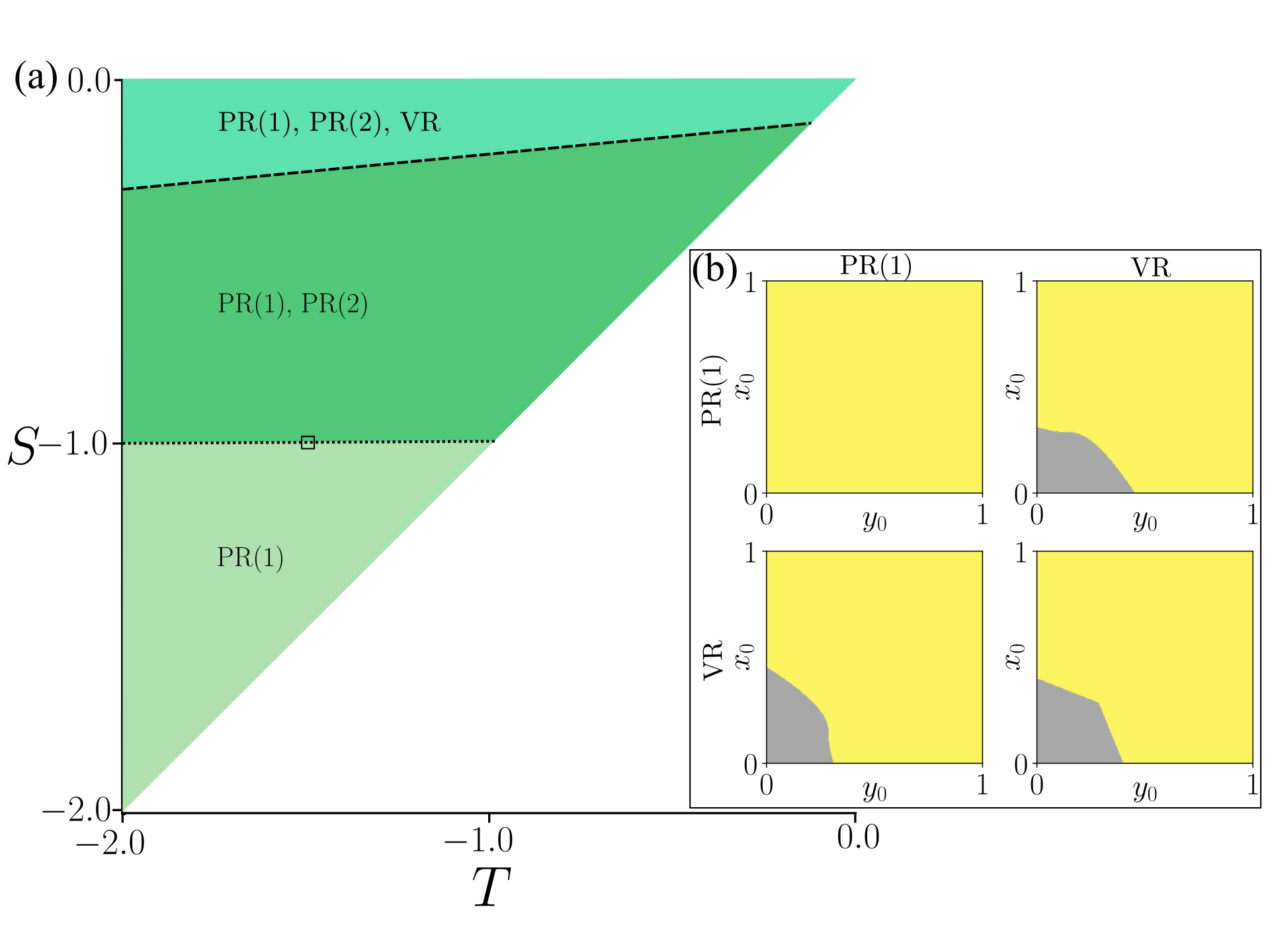}
	\caption{\emph{Evolutionary stability of procedurally rational player}: Sub-figure (a) depicts when {\color{black}PR(1)}, PR(2) and {\color{black}VR} are ESSes. {\color{black}PR(1)} consistently remains an ESS throughout the parameter range corresponding to coordination-II game; it resists the invasion by PR(2) and {\color{black} PR($\infty$), i.e., VR}. While neither PR(2) nor PR($\infty$) is ESS against the intrusion by PR(1) in the region below the dotted line; in the region between the dashed line and the dotted line,  PR(2) (not PR($\infty$)) is an ESS against PR(1). In the region above the dashed line,  PR(2) and PR($\infty$) both are ESSes against PR(1). Sub-figure (b) portrays asymptotic beliefs' dependence on initial beliefs (for the sake of concrete illustration of dependence on initial beliefs, we have fixed parameters $S=-1.0$ and $T=-1.5$---note the small square in sub-figure (a)): abscissas and ordinates, respectively, represent the initial beliefs of player 2 ($y_0$) and player 1 ($x_0$). First and  second rows present player 1 strategies (PR(1) or VR), whereas first and second columns present strategy ({\color{black}PR(1)} or VR) of  player-2. The asymptotic belief of player 1 at each point ($x_0, y_0$) is either $1$ (yellow) or $0$ (grey). }
	\label{fig:c2}
\end{figure}

Given the double averaging involved, it is obvious that the elements of payoff matrix, ${\sf \Pi}$---and hence, if {\color{black}PR(1)} and/or VR is ESS---can be best found numerically. All the numerical codes used to generate the results in the paper are available at Github~\footnote{\url{https://github.com/joydasbairagya/Code-procedurally-rational}.}. The numerical result is best depicted in Fig.~\ref{fig:c2} for coordination-II game: It is crystal clear that a population exclusively consisting of procedurally rational players cannot be invaded by the mutants who are VNM rational. Moreover, for most part (except blue region in the figure) of the $S$-$T$ parameter space, VR is not ESS---a population exclusively consisting of VNM rational players are invaded by the procedurally rational mutants. In conclusion, if there is coordination-II game in a population with only procedurally rational players maintaining cooperation, then any accidental emergence of more rational behaviour (which could have led to population-wide defection) is not amplified and sustained in the population.

In passing, we mention (see Appendices~\ref{app:B},~\ref{app:C}~and~\ref{app:D}) three points highlighting how robust {\color{black}PR(1)} is: Firstly, {\color{black}PR(1)} is a evolutionary stable even if we relax the requirement of infinite population. Assuming that the replication-selection process is Moran-like, and one may define {\color{black}PR(1)} as evolutionary stable in finite population (also called ${\rm ESS}_{N}$~\cite{Nowak2004}) through the following conditions: The fixation probability of VR mutant is less than that due to random drift and the fitness of single VR mutant is less than that of a resident {\color{black}PR(1)}  For coordination-II game, one can show that {\color{black}PR(1)} is ${\rm ESS}_{N}$ for any population size. Secondly, one knows that ESS may not be robust against continuous mutations in the population; hence, finding whether {\color{black}PR(1)} as ESS is stochastically stable~\cite{Foster1990} is very important. We find that while VR is never stochastically stable, {\color{black}PR(1)} is always so. Thirdly, one may wonder what if mutant strategy although VR, adopts a different belief-update scheme, e.g., let's say the belief is updated randomly. Again, even such mutants can't beat {\color{black}PR(1)} in either finite or infinite population.

An immediate extension of the concept of SE(1) is that if each action is independently sampled $k$ times and the action yielding comparatively more cumulative payoff is preferred, then the two procedurally rational players playing with each other reach an equilibrium, termed, SE($k$). It is known that as $k\to\infty$, NE is approached, i.e., $\lim_{k\to\infty}{\rm SE}(k)={\rm NE}$. It implies---since NE corresponds to fully rational player---SE($k+1$) corresponds to procedurally \emph{more} rational player (say, PR($k+1$)-player) than PR($k$)-player (i.e., the one with whom SE($k$) corresponds to). VR-player, in this notation, is PR($\infty$)-player. It is straightforward  (see Appendix~\ref{app:E}) to show that both $(0,0)$ and $(1,1)$ are SE($2$) but unlike PR($1$)-players---but more like  PR($\infty$)-players---both $(0,0)$ and $(1,1)$ are reached under belief-update dynamics depending on initial conditions and parameters. In this backdrop, it is natural to intuit that when pitted against PR($1$), between PR($2$) and PR($\infty$), the former would be ESS over wider range of $S$-$T$ parameter space than the latter; of course, PR($1$) should always be ESS against any invasion by PR($2$) mutants. Indeed, these are what we witness in Fig.~\ref{fig:c2}.

{\color{black}\section{Discussion and Conclusions}
In neoclassical economics, all players playing a game are assumed to be VNM-rational, possessing common knowledge of the game's structure and the players' rationality. These players have been termed VR-player in this paper. Such an idealized player is stylistically called \emph{Homo Economicus}~\cite{Persky1995}---a fictitious being who is popular among researchers because her economic behaviour is somewhat mathematically tractable. The conclusions found for such a player are considered normative. In reality, VR-player is not possible to exist but she exists throughout the research literature of game theory. Many studies of evolution of cooperation are based solely in the context of VR-players; in fact, the well-known paradigmatic game, prisoner's dilemma~\cite{1965_RC}, is a dilemma (resulting in defection when mutual cooperation is better rewarding) only because it is traditionally built with the VR-players. Under such a framework, standing on the shoulders VNM-rationality, both cooperation and defection emerge as possible outcomes (NEs) in the coordination-II game. Consequently, this framework fails to uniquely explain the question why players coordinate on cooperation. 

Our main contribution in this paper is to realize that to address the above question, one must deviate from the assumption of complete rationality. Experimental studies~\cite{Selten2008} of human behaviour have shown that the behaviour aligns more closely with SE($k$) than with NE. Essentially, it is a technical way of saying that real humans---\emph{Homo Sapiens}---are boundedly rational: They do not have full information about the game's components and also do not have infinite capability of calculations to fulfil their selfish goals. This SE($k$) arises when players are procedurally rational, i.e., PR($k$)-players in our notation. This observation naturally leads us to ask: Can procedural rationality account for the emergence of coordinating cooperation? Additionally, we have explored a deeper related question: Is procedural rationality evolutionarily stable, thereby shedding light on the conditions under which bounded rationality persists against completely rational players?  In short, in the backdrop of the canonical problem of  evolution of cooperation in game theory, we are asking who among \emph{Homo Economicus} and \emph{Homo Sapiens} is evolutionarily robust? Our results essentially highlights the possibility that cooperation is seen as an evolutionarily stable feature in the society because real humans are boundedly rational.}

{\color{black}The discovered evolutionary success of procedural rationality in this paper lies in its ability to consistently achieve the payoff dominant coordination, i.e., cooperation; in the coordination-II game, the highest payoff corresponds to cooperation. However, along with cooperation, mutual defection is also a VNM-rational solution for the coordination-II game. Hence, procedural rationality is evolutionarily more effective than VNM-rationality, as the latter cannot yield better payoff. In conclusion, in the simple yet non-trivial setting of $2\times2$ coordination-II games, the reason behind realization of cooperation is the evolutionary stability of procedurally rationality---whether it be in population of any size, whether there be isolated rare mutation to more rational strategies, or whether the mutations be continuous. Thus, it is theoretically clear that \emph{not} being completely (VNM) rational is crucial to allow evolutionary forces resolve the conflict in stag-hunt game kind of scenario in gather-hunter societies of humans, who have evolved to treasure their self-reputation---they would rather be cheated than cheat. This conclusion finds encouragement in the fact that, in general, predicted equilibria under procedural rationality are observed (as in experiments~\cite{Selten2008} with $2\times2$ games) to come closest to observed outcomes. Our work reaffirms the perspective that rather than viewing bounded rationality as a departure from optimality, it should be seen as an evolutionary optimal solution. In passing, we feel a point is worth bringing to the fore explicitly: This paper beautifully sews together concepts from classical game theory~\cite{VNMbook}, behavioural game theory~\cite{camerer2011behavioral} and evolutionary game theory~\cite{smith1982evolution} to arrive at its intriguing results.

One of the main novelties of this paper is that it establishes a framework that challenges the traditional assumption---all players possess the same level of rationality---adopted in most problems solved in the research literature in game theory. While the evolution of different levels of rationality~\cite{Stahl1993} has been studied in the literature, games involving interactions between players with different rationality levels have received little attention. In particular, the concept of different levels of rationality originates from the notion of \textit{BP-rationalizability}~\cite{Bernheim1984}, where rationality is defined in terms of iterative belief-based reasoning under von Neumann--Morgenstern (VNM) expected utility. However, motivated by insights from behavioural economics, we have adopted \emph{procedural rationality}---which better captures bounded human decision-making---as the realistic model of rationality. In this framework, we reinterpret the level of rationality not in terms of VNM rationality, but as the procedural rationality. Naturally, in games where players with different levels of procedural rationality interact, we are able to demonstrate that the resulting equilibrium beliefs differ from known homogeneous solutions.

Additionally, we have constructed a novel evolutionary game-theoretic model in which agents are not genetically hardwired to cooperate or defect. Instead, two interacting players first play an underlying classical game and adopt strategies based on their respective procedural rationality levels. This leads to the emergence of first level of procedural rationality (PR(1)) in gatherer-hunter societies, which in turn fosters cooperation naturally in this social context. Finally, different levels of procedural rationality (PR($k$)) imply different sampling efforts, and hence different cognitive or computational costs. This opens up important avenues for future research into how the cost of information processing, embodied in sampling, shapes the evolution of bounded rationality.}
 
\section*{Data availability statement}
All numerical codes used to generate the data analyzed in this paper are available at Github: \href{https://github.com/joydasbairagya/Code-procedurally-rational}{\url{https://github.com/joydasbairagya/Code-procedurally-rational}}.

\section*{Conflicts of interest statement}
The authors report there are no competing interests to declare.

\begin{acknowledgements}
The authors thank Gregory Kubitz and Jonathan Newton for helpful comments. JDB has been supported by Prime Minister's Research fellowship (govt. of India) and SC acknowledges the support from SERB (DST, govt. of India) through project no. MTR/2021/000119.
\end{acknowledgements}
\appendix
{\color{black}
\section{Mean-field dynamics versus underlying stochastic dynamics}
\label{app:comms}
\begin{figure}[h!]
	\centering
	\includegraphics[width=1.0\linewidth]{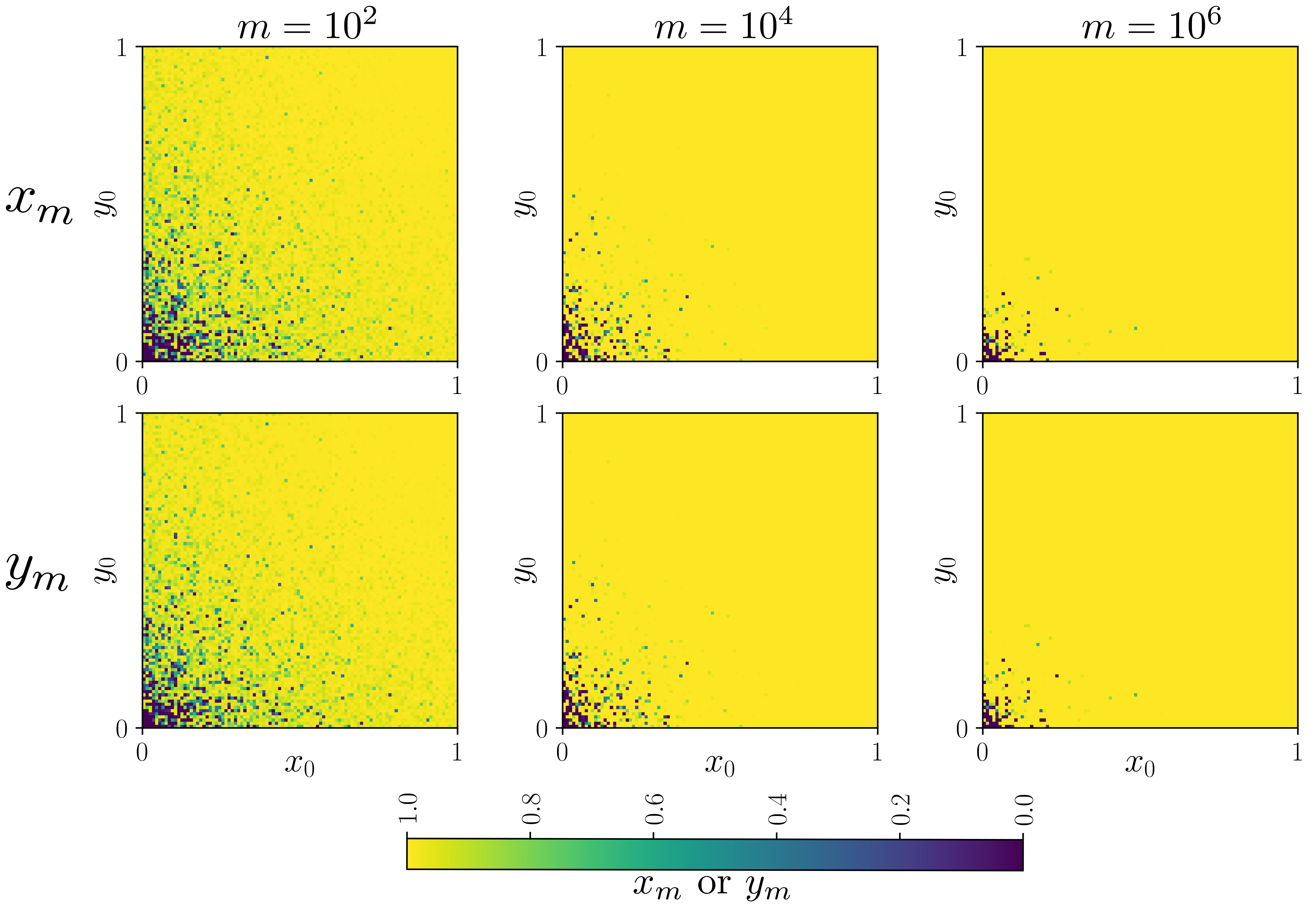}
	\caption{\color{black}$(x_m, y_m)$ consistently deviates away from $(0,0)$ and approaches $(1,1)$ under the stochastic belief update rule. The three columns in the first row show the value of $x_m$ after $m = 10^2$, $10^4$, and $10^6$ steps, respectively, based on belief updating via Eqs.~(\ref{eq:belief_update}). Each point $(x_0, y_0)$ on the subplots is coloured according to the value of $x_m$ obtained after $m$ updates starting from the initial condition $(x_0, y_0)$. Similarly, each point $(x_0, y_0)$ in the subplots of second row is colored based on the value of $y_m$ reached after $m$ updates from that initial condition. Here, $S = -0.5$ and $T = -1.5$. }
	\label{fig:c3}
\end{figure}
Here we numerically demonstrate that for two PR(1)-players, the stochastic belief update rule, Eqs.~(\ref{eq:belief_update}), leads to the same long-term outcome (within expected fluctuations) as the mean-field update equations, Eqs.~(\ref{eq:belief_update_meanfield}). As reported in the main text, the mean-field dynamics for coordination-II games predict that beliefs always diverge from the equilibrium point $(0,0)$ and asymptotically approach $(1,1)$. Consistent with this fact, we observe numerically that the number of initial conditions $(x_0,y_0)$, that evolve toward $(x_m = 1, y_m = 1)$  under the stochastic belief updating rule, increases over time (see Fig.~\ref{fig:c3}). 

To carry out the check, for illustrative purposes, we consider a coordination-II game where $S=-0.5$ and $T=-1.5$. Our conclusion remains unchanged for any other such game. We uniformly sample $100 \times 100$ initial conditions over the range $x_0, y_0 \in [0,1]$, and simulate the belief dynamics independently for $m = 10^2$, $10^4$, and $10^6$ steps using Eqs.~(\ref{eq:belief_update}) to determine the asymptotic $(x_m, y_m)$. Through  Fig.~\ref{fig:c3}, we observe how almost all initial conditions approach $(1,1)$ as time evolves; in fact, even over quite smaller number of iterations ($m$), the lion's share of initial conditions lead to the result same as that of the mean-field dynamics.}

\section{Stability of SE(1)}
\label{app:A}
Here we check stability of sampling equilibria using Eq.~2(a) and~(b) of main text. Before write the equation, we have to know probabilities of best response  ${\rm BR}( \braket{x})=1$, $1/2$ and $0$, given that the corresponding player's belief is $\braket{x}$. Then one can easily calculate the best response $\braket{{\rm BR}(\braket{x})}$. For procedurally rational player, who samples her each action once, ${{\rm BR}(\braket{x})}=\braket{x}+(1-\braket{x})\braket{x}$, for coordination-II game ($1>0>S>T$). Therefore, the evolution equation of mean beliefs follows:
\begin{subequations}
	\label{eq:belief_update_function}
	\begin{eqnarray}
		\frac{d}{d\tau}\braket{x}&=&\braket{y}+(1-\braket{y})\braket{y}-\braket{x},\label{update_ply_1}\\
		\frac{d}{d\tau}\braket{y}&=&\braket{x}+(1-\braket{x})\braket{x}-\braket{y}.\label{update_ply_2}
	\end{eqnarray}
\end{subequations}
 For the above dynamical equation, two fixed points are possible: $(0,0)$ and $(1,1)$. Straightforward linear stability analysis yields that $(0,0)$ is always linearly unstable {(eigenvalues: 1 and $-3$)} and $(1,1)$ is always linearly asymptotically stable {(eigenvalues: $-1$ and $-1$)}.

\section{Finite Population}
\label{app:B}
No population is realistically infinite. In principle, we should analyze the ultimate evolutionary fate of a finite population using a stochastic formalism rather than the deterministic dynamics. A canonical approach to study the evolution of a finite population is the Moran process~\cite{Moran_1958}; accordingly, the concept of the evolutionary stability also modified: ESS modified to ESS$_N$~\cite{Nowak2004}. If the fitness matrix is  ${\sf  \Pi}=\left[^a_c~~~^b_d\right]$, then the conditions for PR to be an  ESS$_N$ are given below (in the limit of weak selection):
\begin{enumerate}
	\item $c(N-1)<b+a(N-2)$,
	\item $d(N-2)+c(2N-1)<b(N+1)+a(2N-4)$;
\end{enumerate}
where $N$ is the fixed size of the population. Now, since ${\sf A}$ corresponds to coordination-II game, we have $a>d$, $b=c$ and $a>b$; consequently, PR is ESS$_N$ for any finite population. 

{\color{black}In passing, in the backdrop of finite population, we remark here that there may be an alternative explanation of achieving equilibrium in  belief other than virtual simulation in mind as discussed in the main text. It is well documented that size of hunter-gatherer societies were often in few dozens~\cite{Bird2019}. Therefore, beliefs about everyone (since there are only a very few individuals) can be updated by directly observing the outcomes of everyday interaction. This way no one has to think what other would do; one only has to update belief from direct observation---the equations for updating  belief still remains same as Eq.~(1) of main text. }
 
\section{Stochastically stable equilibrium (SSE)}
\label{app:C}
In the concept of ESS, small perturbation (e.g., small number of mutants) appears in the resident population only once before it either is wiped out or takes over the population. However, in reality mutants appear incessantly which naturally calls for a stochastic approach of evolutionary stability; one could treat the continuous mutations as a continuos noise which has been modelled as a Gaussian white noise in literature~\cite{Foster1990}. Suppose that fraction of PR-players in an infinite population at time $t$ is $p$ and rest of them are VR-players. Furthermore, if we assume the fitness matrix is ${\sf  \Pi}=\left[^a_c~~~^b_d\right]$, then one can write the time evolution of $p$ as follows:
\begin{equation}
	\frac{dp}{dt}=p(1-p)\left[ap+b(1-p)-cp-d(1-p)\right]+\Gamma(p) \eta(t),
	\label{eq:Langevin}
\end{equation}
where $\eta(t)$ is Gaussian white noise with $\Gamma(p)$ being its strength. To ensure forward invariance, i.e., $p(t)\in [0,1] \forall t$, we have to impose following conditions on $\Gamma(p)$: $\Gamma(p=0)=0$ and $\Gamma(p=1)=0$. Additionally, for analytical tractability, let us assume that $\Gamma(p)=\sigma$ which is independent of $p$. Now, let us examine the steady state probability distribution to see whether there exists any population state ${\bm p^*}\equiv (p^*,(1-p^*))$ such that at $t \to \infty$, the probability of remaining in its neighborhood is non-zero when $\sigma \to 0$. Technically, ${\bm p^*}$ is known as stochastically stable equilibrium (SSE). One could determine the steady state probability distribution from the Fokker--Plank equation corresponding to Eq.~(\ref{eq:Langevin}) with reflecting boundary condition. The steady state probability density $\rho(p)$ comes out to be:
\begin{equation}
	\rho(p)=A \exp\left({\frac{2}{\sigma^2}I(p)}\right),
\end{equation}
where $A$ is a normalization constant and $I(p)\equiv\int p(1-p)\left[ap+b(1-p)-cp-d(1-p)\right] dp =$ \\$\frac{1}{12} p^2 \left[-6 d + 6 b + 4 (2 d + a - 2 b - c) p + 
3 (-d - a + b + c) p^2\right]$. When $\sigma\to0$, the maximum contribution to $\rho(p)$ must come from $p=p^*$ such that $I(p^*)=\max_pI(p)$. It is easy to check that the local maxima of $I(p)$ correspond to $p=0$ and $p=1$; out of these, $p=1=p^*$---the global maximum, and hence the SSE. In conclusion, PR is SSE in any coordination-II game.

\section{Interaction between two players---only one player updates her belief}
\label{app:D} 
The player, who uses rationalizable strategies, opts an action which is a best response according to her belief. However, the belief of the individual, who uses rationalizable strategy, need not to be true strategy of her opponent; instead, it can be random. For mathematical simplicity, let us assume that one player's belief is a realization of a uniform random number between $0$ to $1$: $U[0,1]$. Let us write the belief updating rule for a strategic interaction where Player-1 updates her belief ($x_m$) and Player-2 forms belief ($y_m$) randomly. The belief update naturally should be as follows:
\begin{subequations}
	\label{eq:belief_update_rationalizable}
	\begin{eqnarray}
		{x}_{m+1}&=&\frac{m}{m+1}{x}_{m}+\frac{1}{m+1}{\rm BR}({y}_{m}),\label{update_ply_1}\\
		{y}_{m+1}&=& \xi\sim U[0,1].\label{update_ply_2}
	\end{eqnarray}
	\label{update_ply}
\end{subequations}
It is obvious that $y_m$ can not assume any fixed asymptotic value. 
However, taking average of Eq.~\ref{eq:belief_update_rationalizable}(a), yields: $\braket{x_\infty}=\lim_{m\to\infty}\braket{{\rm BR}(y_{m})}=\braket{{\rm BR}(y_{m})}\,\forall m$. The last equality results from the fact that $y_m$'s are independent uniform random numbers. We also note that $\braket{{\rm BR}(y_{m})}$ is nothing but the probability of playing action C; hence, it is trivially the strategy $q$ of Player-2. Consequently, the asymptotic strategy of Player-1 is $p={\rm BR}(q)$. These conclusions are independent of initial beliefs ($x_0,y_0$).

Let us first consider the case where Player-1 is procedurally rational (PR) and Player-2 is VNM rational (VR$_{\rm ran}$) in a coordination-II game ($1>0>S>T$). Of course, 
\begin{subequations}
	\label{eq:belief_update_function} \label{update_ply_1}
	\begin{eqnarray}
		p&=&{{\rm BR}(q)}=q(2-q),\\
			q&=&{\rm Probability}\left(y_m>\frac{|S|}{1+|T|+|S|}\right)=1-\frac{|S|}{1+|S|+|T|}.\nonumber\\
	\end{eqnarray}
\end{subequations}
The fitness matrix becomes 
\begin{widetext}
\begin{eqnarray}
		{\sf \Pi} = \left[^a_c~~~^b_d\right] =
	\begin{array}{cc|c|c|}
		& \multicolumn{1}{c}{} & \multicolumn{1}{c}{\rm PR} & \multicolumn{1}{c}{\,\,\,\,{\rm VR}_{\rm ran}\,\,\,\,}\\\cline{3-4} 
		\multirow{2}*{{~}} & {\rm PR} & 1 & p[q-|S|(1-q)]-(1-p)q|T|\\\cline{3-4}
		& {\rm VR}_{\rm ran}  & q[p-|S|(1-p)]-(1-q)p|T| &  q[q-(1-q)(|S|+|T|)] \\\cline{3-4} 
			
	\end{array}\,.
    \label{Eq:VR_ran}
\end{eqnarray}
\end{widetext}

A quick inspection of the matrix reveals that PR is the only ESS.

Next consider the case where Player-1 is procedurally rational (PR) and Player-2 is also procedurally rational (PR$_{\rm ran}$). Then, 
\begin{subequations}
	\label{eq:belief_update_function} \label{update_ply_2}
	\begin{eqnarray}
		p&=&{{\rm BR}(q)}=q(2-q),\\
		q&=&\int_{0}^{1}y(2-y) dy.
	\end{eqnarray}
\end{subequations}
Specifically, $(p,q)=(8/9,2/3)$, and hence the fitness matrix (see Eq.~\ref{Eq:VR_ran}) is 
\begin{equation}
	{\sf \Pi} = \left[^a_c~~~^b_d\right] =
	\begin{array}{cc|c|c|}
		& \multicolumn{1}{c}{} & \multicolumn{1}{c}{\rm PR} & \multicolumn{1}{c}{\,\,\,\,{\rm PR}_{\rm ran}\,\,\,\,}\\\cline{3-4} 
		\multirow{2}*{{~}} & {\rm PR} & 1 & \frac{16-8|S|-2|T|}{27}\\\cline{3-4}
		& {\rm PR}_{\rm ran}  &  \frac{16-2|S|-8|T|}{27} &  \frac{4-2|S|-2|T|}{9} \\\cline{3-4}
		
	\end{array}\,.
	\label{Eq:PR_ran}
\end{equation}
Again, only PR is ESS.

Furthermore, in order to extend the robustness of the afore-discussed results to the case of finite populations, let us check if PR is ESS$_N$ as well against the intrusion by VR$_{\rm ran}$ or PR$_{\rm ran}$ mutants. From the fitness matrices (see Eq.~\ref{Eq:VR_ran} and Eq.~\ref{Eq:PR_ran}), one finds that {$a>b>c$ and $a>b>d$}. (To arrive at these, one must use the fact that $p>q$ for the cases in hand.) Consequently, one notes that both the conditions for PR to be ESS$_N$ (see Appendix~\ref{app:B}) are simultaneously satisfied if additionally, $N>2-\frac{3(b-c)}{b-d+2a-2c}$, which anyway is true for all $N\ne 1$. Finally, because of the ordinal relations---{$a>b>c$ and $a>b>d$}---it is trivial to check that PR is SSE against continuous emergence of VR$_{\rm ran}$ and PR$_{\rm ran}$ mutants.

\section{Stability of SE(2) }
\label{app:E}
It is well known if procedurally rational agents sample their each action infinite times in a strategic interaction, then the sampling equilibrium is the Nash equilibrium. However, we have seen that for coordination-II game when agents samples her each action one time, only one sampling equilibrium, $(1,1)$, can be achieved under the sampling dynamics. In the contrary, both pure Nash equilibria can be achieved depending upon the initial belief. We want to examine, at least how many sample one has to do for achieving the other sampling equilibrium, $(0,0)$; we find that if the agent samples her action two times then there are some values of $S$ and $T$ this is achieved. Below we write the  sampling dynamics of a coordination-II when the agents sample their actions two times:
\begin{subequations}
	\label{eq:belief_update_function}
	\begin{eqnarray}
		\frac{d}{d\tau}\braket{x}&=&\braket{y}^2+f(\braket{y})+g(\braket{y})-\braket{x},\label{update_ply_1}\\
		\frac{d}{d\tau}\braket{y}&=&\braket{x}^2+f(\braket{x})+g(\braket{x})-\braket{y}.\label{update_ply_2}
	\end{eqnarray}
\end{subequations}
Here, the form of function $f$ depends on the ordinal relation between  $1+S$ and $0$ and the form of function $g$ depends on the ordinal relation between $2S$ and $T$. Specifically, one has
\begin{enumerate}
	\item for $1+S>0$, $f(z)=2z\left(1-z\right)$,
	\item for $1+S<0$, $f(z)=2z\left(1-z\right)\left[z^2+2z(1-z)\right]$,
	\item for $1+S=0$, $f(z)=2z\left(1-z\right)\left[z^2+2z(1-z)+\frac{1}{2}(1-z)^2\right]$;
\end{enumerate}
and
\begin{enumerate}
	\item for $2S>T$, $g(z)=(1-z)^2\left[z^2+2z(1-z)\right]$,
	\item for $2S<T$, $g(z)=(1-z)^2z^2$, 
	\item for $2S=T$, $g(z)=(1-z)^2\left[z^2+z(1-z)\right]$.
\end{enumerate}
In all the nine possible forms the above dynamical equation, two fixed points  are possible: $(0,0)$ and $(1,1)$. Under linear stability analysis, one easily finds that the fixed pint $(0,0)$ stable if  $f'(0)+g'(0)<1$ and the fixed point $(1,1)$ is stable if $f'(1)+g'(1)<-1$; prime is derivative w.r.t. to the argument of the function. $(1,1)$ is found to be always stable because here $f'(1)+g'(1)=-2$.
%
%

\bibliography{Bairagya}
\end{document}